\documentclass[twocolumn,seceq]{jpsj2}

\def\runauthor{Katsunori \textsc{Kubo} and Yoshio \textsc{Kuramoto}}

\title{
  Octupole Ordering Model for the Phase IV of Ce$_x$La$_{1-x}$B$_6$}

\author{Katsunori \textsc{Kubo}\thanks{E-mail: katukubo@cmpt.phys.tohoku.ac.jp}
  and Yoshio \textsc{Kuramoto}}

\inst{Department of Physics, Tohoku University, Sendai 980-8578}

\recdate{July 2, 2003}

\abst
{
  An octupole ordering model is studied by the mean field theory,
  and its relevance to the phase IV of Ce$_x$La$_{1-x}$B$_6$ is discussed.
  The observed lattice distortion along the [111] direction is interpreted
  in terms of the $\Gamma_{5g}$-type {\it ferro}-quadrupole moment induced by
  an {\it antiferro}-octupole ordered state with $\Gamma_{5u}$ symmetry.
  The octupole model also accounts for the cusp in the magnetization
  as in the N\'{e}el transition,
  and the softening of the elastic constant $C_{44}$ 
  below the ordering temperature.
  However, the internal magnetic field due to the octupole moment
  is smaller than the observed one by an order of magnitude.
  Also discussed is the possibility of a pressure induced antiferromagnetic
  moment in the octupole-ordered state.
}

\kword
{rare-earth hexaboride,
  Ce$_x$La$_{1-x}$B$_6$,
  NdB$_6$, NpO$_2$,
  octupole ordering,
  elastic constant, thermal expansion,
  internal field,
  pressure induced moment}

\begin{document}
\maketitle

\section{Introduction}
Orbital orderings in $f$-electron systems
have attracted much attention.
In usual cases,
an orbital-ordered phase
is a quadrupole-ordered phase.
However,
higher multipole orderings
can also be realized. 
Indeed,
possibilities of octupole orderings
in NpO$_2$~\cite{Santini,Paixao,Caciuffo,Kiss}
and in Ce$_x$La$_{1-x}$B$_6$~\cite{Kusunose,Kuramoto}
have been discussed.
In this paper,
we study
an octupole ordering model for the phase IV of Ce$_x$La$_{1-x}$B$_6$
paying attention
to the coupling between $f$-electrons
and lattice.

CeB$_6$ has a cubic crystal structure
of the CaB$_6$-type,
and Ce ions form a simple cubic lattice.
In this material, there are three phases;
the paramagnetic phase (called phase I),
the antiferro-quadrupole phase (phase II)
and the antiferromagnetic phase (phase III).
The role of the octupole interaction
in these phases
has been
discussed~\cite{Kusunose,Shiina,Sakai,Shiina2}
to explain experimental observations.
In the dilute alloys Ce$_x$La$_{1-x}$B$_6$, 
the so-called phase IV was found 
at $x \simeq 0.75$.~\cite{Sakakibara2}
This phase has attracted much attention.
A large softening of the elastic constant $C_{44}$
just below the transition temperature $T_{\text{I-IV}}$ from the phase I
to the phase IV
was observed.~\cite{Suzuki} 
Furthermore, the magnetic susceptibility shows a cusp 
at $T_{\text{I-IV}}$,
and the magnetization is almost isotropic at ambient pressure.~\cite{Tayama}

It was recently revealed that
the lattice 
shrinks along
the [111] direction in the phase IV.~\cite{Akatsu,Akatsu2}
It is probable that
the softening of $C_{44}$ is due to this distortion.
It may be tempting to ascribe 
the distortion to the ferro $\Gamma_{5g}$-type quadrupole order. 
However, the quadrupole moment is not necessarily the primary order parameter.
Indeed,
the temperature dependence of $C_{44}$ is
different from usual ferro-quadrupole-ordered materials.~\cite{Suzuki}
In the phase I of Ce$_{0.75}$La$_{0.25}$B$_6$,
the temperature dependence of $C_{44}$ follows
the Curie-Weiss law
with an \textit{antiferro-quadrupole} interaction,
but softens drastically just below $T_{\text{I-IV}}$.
The entropy changes more at the transition from the phase I to 
the phase IV than at the other transition
from the phase IV to the phase III.~\cite{Furuno,Suzuki}
On the contrary, 
the entropy changes very little at the quadrupole transition 
in CeB$_6$.~\cite{Furuno}
Thus it is likely that the degeneracy of
each $f$-electron state is already lifted in the phase IV.
This fact also makes it difficult for
the $\Gamma_{5g}$ quadrupoles
to be the primary order parameter in the phase IV.

A large change in the internal field at $T_{\text{I-IV}}$, as probed by
NMR~\cite{Magishi} and $\mu$SR~\cite{Takagiwa}, 
suggests strongly 
that the time reversal symmetry is broken in the phase IV.
A pure quadrupole order
is incompatible with the broken time reversal symmetry.
In addition, a neutron diffraction experiment
found no magnetic reflection in the phase IV.~\cite{Iwasa}
Thus, dipole moments are also unlikely to be the primary order parameter,
although the time reversal symmetry is broken.
Therefore, octupole moments,
which break the time reversal symmetry,
become a candidate for the order parameter
in the phase IV.~\cite{Kuramoto, Kusunose}
In addition,
the order parameter should have an anisotropic nature,
because the cubic symmetry is broken, and the anisotropy
in the magnetization develops
under uniaxial pressure.~\cite{Sakakibara,Sakakibara3, Morie}
Thus the $\Gamma_{5u}$-type octupole moment,
among all octupole moments,
is the most plausible candidate for the order parameter in the phase IV.
We will discuss this point in detail in {\S}~\ref{sec:MF}.

Kusunose and Kuramoto~\cite{Kusunose} have already pointed out
using the Ginzburg-Landau (GL) theory that
the $\Gamma_{5u}$ octupole order
with a finite wave number
should accompany a ferro-quadrupole moment,
and have suggested a possible lattice distortion. 
However, evaluation of the magnitude of the distortion
is beyond their GL theory,
because the GL theory is appropriate
near the transition temperature.
In this paper, we explore in much greater detail
the consequence of the $\Gamma_{5u}$ octupole order
by the mean field theory,
and propose that the lattice distortion
and the softening of the elastic constant $C_{44}$
are due to the order of the $\Gamma_{5u}$ octupole moment.
Some of the results in this paper
were reported in a recent paper.~\cite{Kubo}
In this paper,
we describe the details of the calculation,
and add some new results, especially, the quadrupole susceptibilities.

This paper is organized as follows.
In {\S}~\ref{sec:MF},
we introduce an octupole ordering model
and solve this model by using the mean field theory.
In {\S}~\ref{sec:M}, {\S}~\ref{sec:Ela} and {\S}~\ref{sec:Lattice},
we discuss the magnetization, elastic constants
and lattice distortion, respectively, in the octupole-ordered state.
In {\S}~\ref{sec:field},
we estimate the internal field
from the octupole moment.
In {\S}~\ref{sec:pressure},
we discuss a possible dipole moment
under uniaxial pressure in the octupole-ordered state.
The last section is devoted to a summary.

\section{Mean Field Theory}\label{sec:MF}
The crystalline electric field (CEF)
ground state of
Ce$^{3+}$ ($J=5/2$) in
Ce$_x$La$_{1-x}$B$_6$ is
the $\Gamma_8$ quartet.~\cite{Zirngiebl,Luthi,Sato}
The excited level $\Gamma_7$ lies about 500~K
from the $\Gamma_8$ level
and is neglected.
The $\Gamma_8$ states are represented in terms of eigenstates of $J_z$ as
\begin{align}
  |+ \uparrow \rangle
  &= \sqrt{5/6}|+5/2 \rangle
  +\sqrt{1/6}|-3/2 \rangle,\label{eq:base1}\\
  |+ \downarrow \rangle
  &= \sqrt{5/6}|-5/2 \rangle
  +\sqrt{1/6}|+3/2 \rangle,\\
  |- \uparrow \rangle 
  &= |+1/2 \rangle,\\
  |- \downarrow \rangle 
  &= |-1/2 \rangle.\label{eq:base2}
\end{align}
Within the $\Gamma_8$ quartet,
the number of independent multipolar moments is 15,
and active octupole moments have either $\Gamma_{2u}$, $\Gamma_{4u}$
or $\Gamma_{5u}$ symmetry.~\cite{Shiina}

We first discuss the properties of the
$\Gamma_{2u}$ and $\Gamma_{4u}$ moments.
The $\Gamma_{4u}$-type octupole moments
accompany dipole moments,
because dipole moments have the same symmetry $\Gamma_{4u}$.~\cite{Shiina}
Thus the $\Gamma_{4u}$ octupole moments
are unlikely to be the order parameter in the phase IV.
The $\Gamma_{2u}$-type octupole moment,
as proposed in ref.~\citen{Kuramoto},
is pseudoscalar, and
eigenstates of the $\Gamma_{2u}$ moment
have the cubic symmetry.
Hence it does not accompany quadrupole moments,
and it seems difficult to explain
the lattice distortion in the phase IV
by the $\Gamma_{2u}$ octupole orderings.
In the followings,
we study the properties of $\Gamma_{5u}$ octupole moments,
and discuss their possibility as the order parameter
of the phase IV.

Let us introduce pseudospins
$\mib{\sigma}$ and $\mib{\tau}$ to describe the $\Gamma_8$ quartet:
\begin{equation}
  \tau^z|\pm \sigma^z \rangle = \pm|\pm \sigma^z \rangle,
\end{equation}
\begin{equation}
  \sigma^z|\tau^z \uparrow \rangle = +|\tau^z \uparrow \rangle, \ 
  \sigma^z|\tau^z \downarrow \rangle = -|\tau^z \downarrow \rangle,
\end{equation}
and the transverse components which flip the pseudospins.
The relevant multipole
moments are given
with the notation $(\alpha,\beta,\gamma)=(x,y,z)$, $(y,z,x)$ or $(z,x,y)$
by
\begin{align}
  \mib{J} &=
  \frac{7}{6}\left[\mib{\sigma}
    +\frac{4}{7}(\eta^+ \sigma_x,\eta^- \sigma_y,\tau_z \sigma_z)
    \right],\\
  O^0_2 &=
  \frac{1}{\sqrt{3}}(2J^2_z-J^2_x-J^2_y)
  =\frac{8}{\sqrt{3}}\tau^z,\label{eq:O20}\\
  O^2_2 &=J^2_x-J^2_y=\frac{8}{\sqrt{3}}\tau^x,\\
  O_{\beta \gamma} &=2 \overline{J_\beta J_\gamma}
  =\frac{2}{\sqrt{3}}\tau^y \sigma^\alpha,\label{eq:Obc}\\
  T^{5u}_\alpha &=
  \frac{1}{2\sqrt{3}}
  (\overline{J_\alpha J^2_\beta}-\overline{J^2_\gamma J_\alpha}),
\end{align}
and
\begin{equation}
  (T^{5u}_x,T^{5u}_y,T^{5u}_z)
  =(\zeta^+ \sigma^x,\zeta^- \sigma^y,\tau^x \sigma^z),
\end{equation}
where bars on the products represent normalized symmetrization,
\textit{e.g.}, $\overline{J_xJ^2_y} =(J_xJ^2_y+J_yJ_xJ_y+J^2_yJ_x)/3$, 
and we have introduced the notations
\begin{align}
  \eta^{\pm}&=\frac{1}{2}(\pm \sqrt{3} \tau^x-\tau^z ),\\
  \zeta^{\pm}&=-\frac{1}{2}(\tau^x \pm \sqrt{3} \tau^z).
\end{align}

\subsection{Level Scheme}
Before introducing a octupole ordering model on a lattice,
we discuss the `easy axis' of the $\Gamma_{5u}$ octupole moment
under a fictitious octupole field
$\mib{A}^{5u} = (A^{5u}_x,A^{5u}_y,A^{5u}_z)$. 
It is sufficient to consider
high-symmetry directions of $\mib{A}^{5u}$.
With $|\mib{A}^{5u}|=1$, the four eigenvalues (Ev) of relevant cases
are given as follows:\\
(i) $\mib{A}^{5u} \parallel (0,0,1)$:
Ev$[T^{5u}_z]=\pm 1$ (both doubly degenerate);\\
(ii) $\mib{A}^{5u} \parallel (1,1,0)$:
Ev$[(T^{5u}_x+T^{5u}_y)/\sqrt{2}]=\pm (\sqrt{3} \pm 1)/2$;\\
(iii) $\mib{A}^{5u} \parallel (1,1,1)$
: Ev$[(T^{5u}_x+T^{5u}_y+T^{5u}_z)/\sqrt{3}]=\pm \sqrt{2}$
and 0 (doubly degenerate).\\
Case (iii) gives the largest eigenvalue,
and thus the easy axis
is along the [111] direction
and equivalent ones.

\begin{figure}[t]
  \includegraphics[width=8cm]{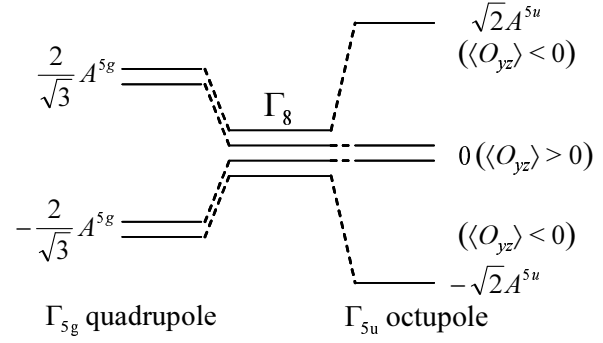}
  \caption{The level scheme in the $\Gamma_{5u}$
    interaction (right) and in the $\Gamma_{5g}$ interaction (left).
    The quantities $A^{5u}$ and $A^{5g}$
    are the octupolar and quadrupolar mean fields.}
  \label{figure:level}
\end{figure}

In the mean field theory, the fictitious octupole field represents the
intersite interaction.  
When the intersite interaction is isotropic, a state with
$(|\langle T^{5u}_x \rangle |,
|\langle T^{5u}_y \rangle |,
|\langle T^{5u}_z \rangle |) \parallel (1,1,1)$
is the most stable.
We note that the operator
$T^{5u}_x + T^{5u}_y + T^{5u}_z$
commutes with
$O_{yz}+O_{zx}+O_{xy}$,
and
these two operators can be diagonalized simultaneously.
On the right side of Fig.~\ref{figure:level},
we illustrate the level scheme
in octupole field $\mib{A}^{5u}=A^{5u}(1,1,1)/\sqrt{3}$.
The $\Gamma_8$ level splits into
three levels.
Not only the time reversal symmetry,
but also the cubic symmetry is broken.
The multipolar moments in
the ground and highest
states are
\begin{align}
  \langle T^{5u}_x \rangle=\langle T^{5u}_y \rangle
  &=\langle T^{5u}_z \rangle=\mp \sqrt{2/3},\label{eq:T_a}\\
  \langle O_{yz} \rangle=\langle O_{zx} \rangle
  &=\langle O_{xy} \rangle=-2/3,\label{eq:O_ab}  
\end{align}
and the others are zero.
Thus,
ferro, antiferro and other collinearly
ordered states with
$(\langle T^{5u}_x \rangle,
  \langle T^{5u}_y \rangle,
  \langle T^{5u}_z \rangle ) \parallel (1,1,1)$
have a homogeneous $\Gamma_{5g}$ moment.
With the quadrupole-strain coupling, the crystal should distort along  [111].

It is obvious that a $\Gamma_{5g}$-type ferro-quadrupole interaction alone
can lead to
$\langle O_{yz} \rangle =\langle O_{zx} \rangle =\langle O_{xy} \rangle \ne 0$.
The Kramers degeneracy remains in this case.
The energy level splits into two levels,
and the ground state is twofold degenerate as shown on the left side of 
Fig.~\ref{figure:level}.

\subsection{Antiferro-Octupole Ordering Model}
We proceed to consider
the $\Gamma_8$ quartets
on a simple cubic lattice.
From the above discussion,
we find that 
the order parameter which
breaks the time reversal symmetry 
and induces quadrupole moments,
but which accompanies no dipole moment
is only the $\Gamma_{5u}$ octupole moment.
However, we cannot determine
the periodicity of the $\Gamma_{5u}$-ordered state in the phase IV
only from the above consideration,
because any collinearly ordered state with
$(\langle T^{5u}_x \rangle,
  \langle T^{5u}_y \rangle,
  \langle T^{5u}_z \rangle ) \parallel (1,1,1)$
has a uniform $\Gamma_{5g}$ moment.
In the present study,
we consider a G-type (staggered) antiferro-octupole order
as the simplest example of $\Gamma_{5u}$ orders.
The importance of the $\Gamma_{5u}$ nearest-neighbor interaction
in causing the change from the phase III to the phase III$^{\prime}$,
even with a weak magnetic field, 
was pointed out in ref.~\citen{Kusunose}.
We consider only this nearest-neighbor interaction, and
take the following model:
\begin{equation}
  \mathcal{H}=I^{5u} \sum_{(i,j)}
  \sum_{\alpha=x,y,z}T^{5u}_{\alpha \ i} T^{5u}_{\alpha \ j},
\end{equation}
where $(i,j)$ denotes a nearest-neighbor pair.
We study this Hamiltonian by the mean field theory, and
choose the value of $I^{5u}$ so as
to reproduce the transition temperature
$T_{\text{I-IV}}$ in Ce$_{0.75}$La$_{0.25}$B$_6$,
i.e., $T_{5u}=6I^{5u}=1.7$~K.
In carrying out the mean field theory,
we assume the G-type two sublattice structure,
but do not assume a particular solution, \textit{e.g.},
$\langle T^{5u}_x \rangle
=\langle T^{5u}_y \rangle
=\langle T^{5u}_z \rangle$.
When we obtain several self-consistent solutions,
we choose the one which has
the lowest free energy.

\begin{figure}[t]
  \includegraphics[width=8cm]{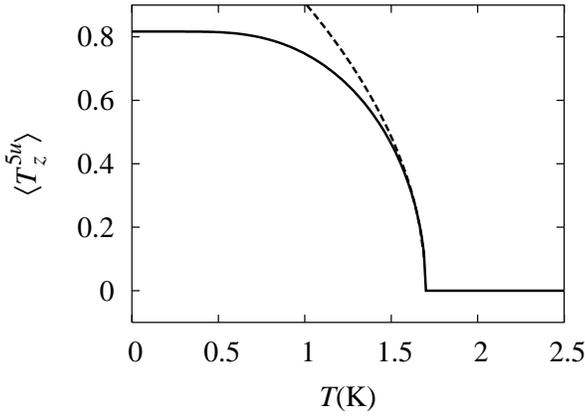}
  \caption{Temperature dependence of
    the antiferro-octupole moment
    $\langle T^{5u}_z \rangle$
    $(=\langle T^{5u}_x \rangle =\langle T^{5u}_y \rangle )$.
    The dashed line is
    the asymptotic behavior $\sqrt{2}(1-T/T_{5u})^{1/2}$
    at $T \simeq T_{5u}$.
    }
  \label{figure:Tbx}
\end{figure}

\begin{figure}[t]
  \includegraphics[width=8cm]{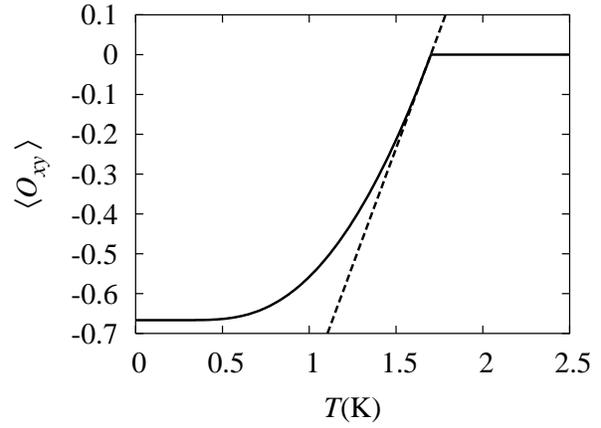}
  \caption{Temperature dependence of
    the ferro-quadrupole moment
    $\langle O_{xy} \rangle$
    $(=\langle O_{yz} \rangle =\langle O_{zx} \rangle )$.
    The dashed line is
    the asymptotic behavior $-2(1-T/T_{5u})$ at $T \simeq T_{5u}$.
    }
  \label{figure:Oyz}
\end{figure}

The temperature dependences of
$\langle T^{5u}_z \rangle$ and $\langle O_{xy} \rangle$
are shown in Figs.~\ref{figure:Tbx} and ~\ref{figure:Oyz}.
The solution obtained has the antiferro-octupole order with
$\langle T^{5u}_x \rangle=\langle T^{5u}_y \rangle =\langle T^{5u}_z \rangle$,
accompanying the ferro-quadrupole moment
$\langle O_{yz} \rangle =\langle O_{zx} \rangle =\langle O_{xy} \rangle$.
At $T \simeq T_{5u}$, we obtain
$\langle T^{5u}_z \rangle \propto (1-T/T_{5u})^{1/2}$
as a consequence of the mean field theory.
On the other hand,
the quadrupole moment behaves as
$\langle O_{xy} \rangle \propto (1-T/T_{5u})$,
which indicates the feature of the induced order parameter.~\cite{Kusunose}
We also find other equivalent solutions:
$(\langle T^{5u}_\alpha \rangle,
\langle T^{5u}_\beta \rangle,
\langle T^{5u}_\gamma\rangle,
\langle O_{\beta \gamma}\rangle,
\langle O_{\gamma \alpha}\rangle,
\langle O_{\alpha \beta}\rangle)
= (\pm B,\pm B,\mp B,+C,+C,-C)$
with
$(\alpha,\beta,\gamma)=(x,y,z)$, $(y,z,x)$ or $(z,x,y)$,
where quantities $B$ and $C$
are positive and
depend on temperature.
The degeneracy of these solutions
can be lifted by a magnetic field
or pressure.
In the following sections,
we choose a domain in which the free energy is minimized.

\section{Magnetization}\label{sec:M}
A magnetic field can lift the degeneracy of the domains.
For $\mib{H} \parallel [111]$,
the (111) domain given by eqs.~(\ref{eq:T_a}) and (\ref{eq:O_ab})
is not the most stable one.
Instead, 
a state where
two of $\langle O_{\alpha \beta}\rangle$'s have positive values,
and the other has a negative value is stabilized.
For $\mib{H} \parallel [110]$,
$\langle J_x \rangle$ and $\langle J_y \rangle$
have positive values,
and a state with $\langle O_{xy} \rangle>0$ is realized.
On the other hand,
all the domains obtained in {\S}~\ref{sec:MF} have the same energy
under $\mib{H} \parallel [001]$.

\begin{figure}[t]
  \includegraphics[width=8cm]{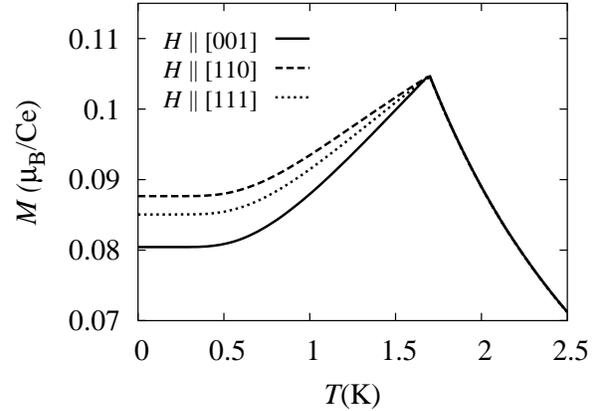}
  \caption{Temperature dependence of the magnetization
    in magnetic field $H=0.2$~T along various directions.
    The domains are selected by the magnetic field directions.
  }
  \label{figure:M}
\end{figure}

\begin{figure}[t]
  \includegraphics[width=8cm]{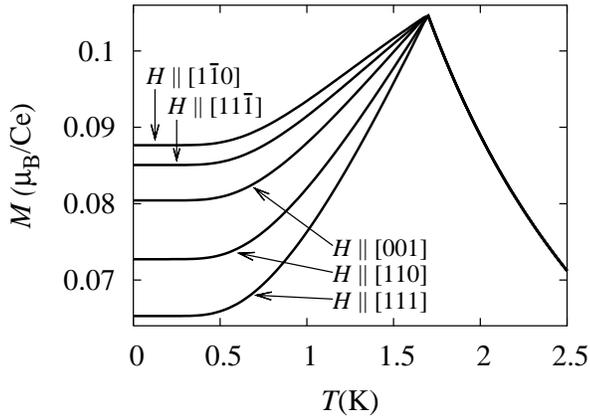}
  \caption{Temperature dependence of the magnetization
    in the single domain with
    $\langle T^{5u}_x \rangle = \langle T^{5u}_y \rangle
    = \langle T^{5u}_z \rangle$
    in magnetic field $H=0.2$~T along various directions.
  }
  \label{figure:single-domain}
\end{figure}

In Fig.~\ref{figure:M},
we show the temperature dependence of
the magnetization
in magnetic field $H=0.2$~T
along the three high-symmetry directions.
The magnetization has a cusp at the transition temperature $T_{5u}$.
The presence of a cusp is consistent 
with experimental observations.~\cite{Tayama,Sakakibara,Sakakibara3, Morie}
Very recently, magnetization
under uniaxial pressure $p=100$~MPa along the [111] direction
has been measured at $H=0.2$~T and $H=0.4$~T.~\cite{Morie}
The observed magnetization is anisotropic
and $M_{[1\bar{1}0]} > M_{[001]} > M_{[111]}$.
It seems that the system under the uniaxial pressure has a single domain.
In order to deal with this situation,
we have computed the temperature dependence of
the magnetization assuming a single domain with
$\langle T^{5u}_x \rangle = \langle T^{5u}_y \rangle
= \langle T^{5u}_z \rangle$
($\langle O_{yz} \rangle=\langle O_{zx} \rangle=\langle O_{xy} \rangle$)
in magnetic field $H=0.2$~T.
The results are shown in Fig.~\ref{figure:single-domain}.
The anisotropy derived by our model is consistent
with the experimental observation.

On the other hand, a previous experiment without an external stress,
the magnetization below $T_{\text{I-IV}}$ is almost isotropic.~\cite{Tayama}
The observed isotropy may be due to
a multi-domain structure.
Our model have four equivalent solutions
in the absence of a magnetic field.
If these four domains are distributed with the same probability,
the magnetic susceptibility is isotropic,
because the cubic symmetry is effectively recovered in the multi-domain state.
The anisotropy in the magnetization under $p \parallel [001]$
was also reported.~\cite{Sakakibara,Sakakibara3}
It is not easy to explain this anisotropy,
because the four domains in our model are equivalent under $p \parallel [001]$
and it seems that a single domain state is not realized
under $p \parallel [001]$.

We have also calculated
the magnetization in the case of ferro-octupole interaction.
The result is qualitatively similar to that in the antiferro-octupole case;
the magnetization has a cusp,
and the anisotropy
in the single domain with
$\langle T^{5u}_x \rangle = \langle T^{5u}_y \rangle
= \langle T^{5u}_z \rangle$
is the same as that of the antiferro-octupole case.
Therefore,
it is difficult to exclude the possibility of
the ferro-octupole ordering in the phase IV
from the magnetization alone.

\begin{figure}[t]
  \includegraphics[width=8cm]{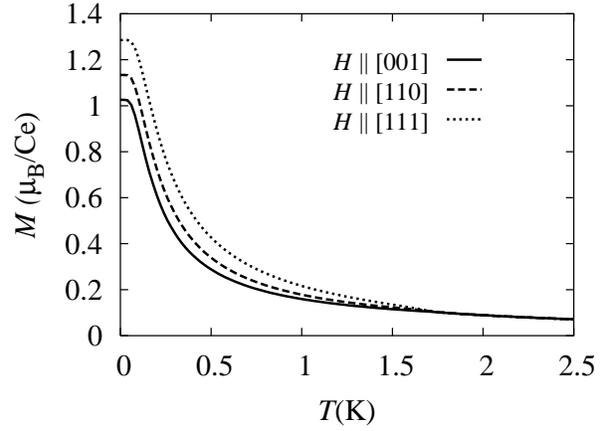}
  \caption{Temperature dependence of the magnetization
    in magnetic field $H=0.2$~T along various directions
    for the case where the $\Gamma_{5g}$ quadrupole moment is the primary
    order parameter.}
  \label{figure:qua5m}
\end{figure}

For comparison,
we show
the magnetization of a model
in which 
only the $\Gamma_{5g}$ ferro-quadrupole interaction is present
in Fig.~\ref{figure:qua5m}.
The interaction between quadrupole moments
is chosen so as to reproduce the transition temperature
$T_{\text{I-IV}}=1.7$~K.
In this calculation,
the domains are determined by the direction of a magnetic field.
We cannot fix a domain in this case,
because the ground state is continuously degenerate
in the absence of a magnetic field.
For example,
the state with
$\langle O_{yz} \rangle=\langle O_{zx} \rangle=\langle O_{xy} \rangle$
is not even metastable in $\mib{H} \parallel [001]$.
The magnetization only slightly changes  
at the quadrupole transition temperature, 
and does not have a cusp.
Thus it is unlikely that
the primary order parameter in the phase IV
is the ferro-quadrupole moment.

\section{Elastic Constant}\label{sec:Ela}
In ferro-quadrupole-ordered materials such as TmZn~\cite{Morin},
the elastic constant often
shows a large softening, following the Curie-Weiss law, 
as the transition temperature is approached from above.
However,
the softening of $C_{44}$ above $T_{\text{I-IV}}$
is very small
in Ce$_{0.75}$La$_{0.25}$B$_{6}$,
and a large softening is observed
just below $T_{\text{I-IV}}$.~\cite{Suzuki}
Moreover, an
antiferro-quadrupole interaction
$g^{\prime}_{\Gamma_5}=-2$~K
is deduced by a fitting of
the temperature dependence of $C_{44}$ above $T_{\text{I-IV}}$.
Thus it is difficult to ascribe this softening
to a ferro-quadrupole interaction.
In this section,
we consider the elastic constant
in the octupole ordering model.
We do not include the quadrupole-quadrupole interaction
$g^{\prime}_{\Gamma_5}$ for simplicity.

\begin{figure}[t]
  \includegraphics[width=8cm]{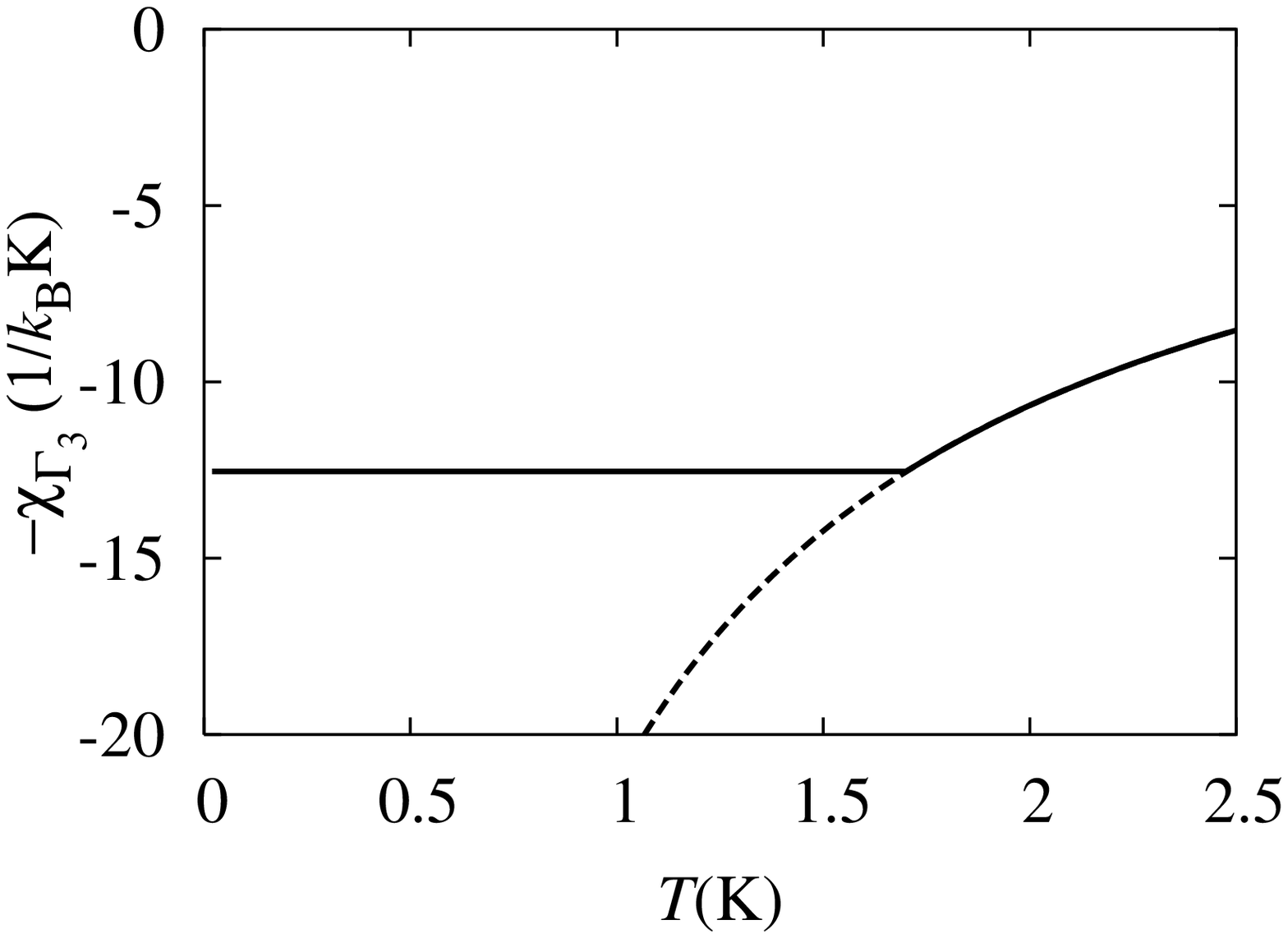}
  \caption{Temperature dependence of the
    quadrupole susceptibility $\chi_{\Gamma_3}$ (solid line).
    The Curie law behavior
    $\chi^{\text{C}}_{\Gamma_3}=64/(3k_{\text{B}}T)$
    at high temperatures is also shown (dashed line).}
  \label{figure:chi3}
\end{figure}

\begin{figure}[t]
  \includegraphics[width=8cm]{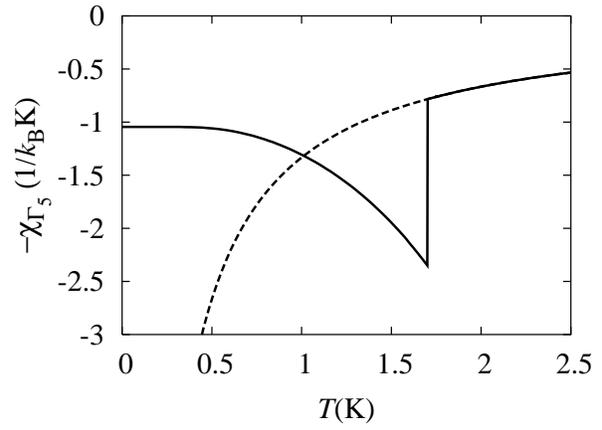}
  \caption{Temperature dependence of the
    quadrupole susceptibility $\chi_{\Gamma_5}$ (solid line).
    The Curie law behavior
    $\chi^{\text{C}}_{\Gamma_5}=4/(3k_{\text{B}}T)$
    at high temperatures is also shown (dashed line).}
  \label{figure:chi5}
\end{figure}

The elastic energy associated with the quadrupole moments is given 
per unit volume by
\begin{equation}
  E = \sum_{\Gamma \gamma}
  \left(
    \frac{1}{2}\epsilon^2_{\Gamma \gamma} C^{(0)}_{\Gamma}
    +g_{\Gamma} \sum_{i}
    \epsilon_{\Gamma \gamma} O_{\Gamma \gamma i}
  \right),\label{eq:Eela}
\end{equation}
where 
$\epsilon_{\Gamma \gamma}$ is the strain tensor,
$g_{\Gamma}$ is the magneto-elastic coupling constant, and
$C^{(0)}_{\Gamma}$ is the bare elastic constant.
The second sum runs over Ce sites $i$ in a unit volume.
The second derivative of the free energy
with respect to $\epsilon_{\Gamma \gamma}$
gives the elastic constant $C_{\Gamma}$,~\cite{Callen}
which is given by
\begin{equation}
  C_{\Gamma}=C^{(0)}_{\Gamma}
  -N g^2_{\Gamma}\chi_{\Gamma},
\end{equation}
where $N$ is the number of Ce ions in the unit volume
and $\chi_{\Gamma}$ is the quadrupole susceptibility
defined by
\begin{equation}
  \begin{split}
    \chi_{\Gamma}
    =&-\frac{\text{d} \langle O_{\Gamma \gamma} \rangle}
    {\text{d} (g_{\Gamma}\epsilon_{\Gamma \gamma})}
    \Biggr|_{g_{\Gamma}\epsilon_{\Gamma \gamma}=0}\\
    =&
    \frac{1}{k_{\text{B}}T}
    \sum_n p_n
    \left[
      \langle n|O_{\Gamma \gamma}|n \rangle
      -\langle O_{\Gamma \gamma} \rangle_0
    \right]
    \frac{\text{d} E_n}{\text{d} (g_{\Gamma}\epsilon_{\Gamma \gamma})}
    \Biggr|_{g_{\Gamma}\epsilon_{\Gamma \gamma}=0}\\
    &-\sum_n p_n
    \frac{\text{d} \langle n|O_{\Gamma \gamma}|n \rangle}
    {\text{d} (g_{\Gamma}\epsilon_{\Gamma \gamma})}
    \Biggr|_{g_{\Gamma}\epsilon_{\Gamma \gamma}=0},
  \end{split}
  \label{eq:chiG}
\end{equation}
where $E_n$ and $|n\rangle$
are the eigenvalue and eigenstate of the total Hamiltonian,
and $\langle \cdots \rangle_0$ denotes
the expectation value without the strain.
The occupation probability $p_n$ is given by
\begin{equation}
  p_n=\frac{e^{-E_n/k_{\text{B}}T}}
  {\sum_m e^{-E_m/k_{\text{B}}T}}.
\end{equation}
The elastic constants,
strain tensors
and quadrupole moments
with $\Gamma_3$ and $\Gamma_5$ symmetries
are given as follows:
\begin{align}
  C_{\Gamma_3} &= (C_{11}-C_{12})/2,\\
  C_{\Gamma_5} &= 4C_{44},\\
  \epsilon_{\Gamma_3 u} &= \epsilon_u=
  (2\epsilon_{zz}-\epsilon_{xx}-\epsilon_{yy})/\sqrt{3},\\
  \epsilon_{\Gamma_3 v} &=\epsilon_v=\epsilon_{xx}-\epsilon_{yy},\\
  \epsilon_{\Gamma_5 \alpha \beta} &= \epsilon_{\alpha \beta},\\
  O_{\Gamma_3 u} &= O^0_2,\\
  O_{\Gamma_3 v} &= O^2_2,\\
  O_{\Gamma_5 \alpha \beta} &= O_{\alpha \beta}.
\end{align}
We calculate $\chi_{\Gamma}$
by differentiating $\langle O_{\Gamma \gamma} \rangle$
with respect to $g_{\Gamma}\epsilon_{\Gamma \gamma}$
numerically.
In Figs.~\ref{figure:chi3} and \ref{figure:chi5},
we show
the temperature dependence of
$\chi_{\Gamma_3}$ and $\chi_{\Gamma_5}$.
Above the transition temperature $T_{5u}$,
the octupole interaction $I^{5u}$ does
not influence
$\chi_{\Gamma_3}$ and $\chi_{\Gamma_5}$,
because the strain $\epsilon_{\Gamma \gamma}$
does not induce octupole moments
in the non-ordered state.
The quadrupole susceptibility $\chi_{\Gamma_5}$
jumps at $T_{5u}$
due to the octupole order.
Although the mean field theory gives a finite jump at $T_{5u}$,
fluctuations beyond the mean field theory should lead to divergence of 
$\chi_{\Gamma_5}$ at $T_{5u}$.

For further understanding of the jump obtained above,
we explore $\chi_{\Gamma_5}$ at $T \simeq T_{5u}$
analytically.
The mean field Hamiltonian
for the A-sublattice
with finite $\epsilon_{xy}$
without magnetic field
is given by
\begin{equation}
  \mathcal{H}_{\text{MF}}
  =-6I^{5u}
  \sum_{\alpha}\langle T^{5u}_{\alpha} \rangle T^{5u}_{\alpha}
  +g_{\Gamma_5} \epsilon_{xy} O_{xy}.
\end{equation}
For $T<T_{5u}$ and $\epsilon_{xy} \ne 0$,
we have $\langle T^{5u}_x \rangle
=\langle T^{5u}_y \rangle \ne \langle T^{5u}_z \rangle$.
We determine these quantities self-consistently.
We obtain $E_n$ and $|n\rangle$
of $\mathcal{H}_{\text{MF}}$
by perturbation theory
and calculate $\chi_{\Gamma_5}$
by using eq.~(\ref{eq:chiG}).
At $T \simeq T_{5u}$,
we obtain
\begin{equation}
  \chi_{\Gamma_5}=
  \frac{4}{3k_{\text{B}}T}
  -\frac{2\sqrt{6}I^{5u}}{k_{\text{B}}T}
  \chi^{5u\text{-}xy}_x h,
  \label{eq:chiRPA}
\end{equation}
where
\begin{equation}
  h=-\frac{E_g}{k_{\text{B}}T}
  =\frac{2\sqrt{6}I^{5u}}{k_{\text{B}}T}
  \langle T^{5u}_x+T^{5u}_y+T^{5u}_z \rangle_0,
\end{equation}
with $E_g$  the mean field ground state energy
per Ce ion.
We have also introduced
the octupole susceptibility
with respect to the strain $\epsilon_{xy}$,
which is defined by
\begin{equation}
  \chi^{5u\text{-}xy}_{\alpha}
  =-\frac{\text{d} \langle T^{5u}_{\alpha} \rangle}
  {\text{d} (g_{\Gamma_5}\epsilon_{xy})}
  \Biggr|_{g_{\Gamma_5}\epsilon_{xy}=0}.
\end{equation}
By symmetry, we obtain
\begin{equation}
  \chi^{5u\text{-}xy}_{x}=\chi^{5u\text{-}xy}_{y}.
\end{equation}
At $T > T_{5u}$,  we have 
$h=0$ and $\chi^{5u\text{-}xy}_{\alpha}=0$,
and then only
the first term in eq.~(\ref{eq:chiRPA}),
i.e, the Curie term remains finite.
The second term in eq.~(\ref{eq:chiRPA})
becomes finite at $T < T_{5u}$.
The octupole susceptibility satisfies the equation
\begin{equation}
  h^2
  \begin{pmatrix}
    2 & 1 \\ 1 & 5
  \end{pmatrix}
  \begin{pmatrix}
    \chi^{5u\text{-}xy}_z \\ \chi^{5u\text{-}xy}_x
  \end{pmatrix}
  =
  -\frac{6\sqrt{6}}{k_{\text{B}}T} h
  \begin{pmatrix}
    0 \\ 1
  \end{pmatrix}
  .
\end{equation}
When $h \ne 0$, i.e,
$T < T_{5u}$,
we obtain
\begin{align}
  \chi^{5u\text{-}xy}_z
  &=-\frac{1}{2} \chi^{5u\text{-}xy}_x,\\
  \chi^{5u\text{-}xy}_x
  &=-\frac{1}{k_{\text{B}}T}
  \frac{4\sqrt{6}}{3}h^{-1}.
\end{align}
Therefore
$\chi^{5u\text{-}xy}_x$
diverges at $T_{\text{I-IV}}$,
and the second term in eq.~(\ref{eq:chiRPA})
becomes finite just below $T_{5u}$.
The change of the susceptibility $\Delta \chi_{\Gamma_5}$
at $T_{5u}$ is given by
\begin{equation}
  \Delta \chi_{\Gamma_5}
  =\frac{8}{3k_{\text{B}}T_{5u}}.
\end{equation}

Experimentally, $C_{44}$ 
remains finite at $T_{\text{I-IV}}$.
Moreover,
$C_{44}$ becomes smaller as temperature is lowered below $T_{\text{I-IV}}$.
In general,
the second-order transition involving the lattice distortion along [111]
leads to divergence of $C_{44}$ at the transition temperature.
Thus the observed behavior of the elastic constant seems
to be influenced by the experimental setup,
and may not be intrinsic to $C_{44}$.
One possible reason for this behavior 
is that the large ultrasonic absorption
accompanying softening makes it difficult
to determine the elastic constant accurately.
Another possibility is the mode mixing of the sound wave.
To determine $C_{44}$,
sound wave with wave vector
$\mib{k} \parallel [001]$ and polarization $\mib{u} \parallel [100]$
was used in the experiment.~\cite{Suzuki}
However, this mode is not a normal mode below $T_{\text{I-IV}}$,
because the cubic symmetry is broken.
It is desirable to measure the longitudinal ultrasound velocity
with $\mib{k} \parallel [111]$ 
which is expected to reflect the anomaly in $C_{44}$ directly,
because this mode remains a normal mode even in the trigonal symmetry
($\langle O_{yz} \rangle
= \langle O_{zx} \rangle
= \langle O_{xy} \rangle \ne 0$).

\section{Lattice Distortion}\label{sec:Lattice}
We now discuss the lattice distortion
due to the antiferro-octupole order.
By minimizing the elastic energy (\ref{eq:Eela})
associated with the $\Gamma_{5g}$ moments,
we obtain
\begin{equation}
  \epsilon_{\alpha \beta} =-\frac{g_{\Gamma_5}}{4 C^{(0)}_{44}}
  \sum_{i} \langle O_{\alpha \beta \ i} \rangle.
\end{equation}
We use the following experimental values for Ce$_{0.75}$La$_{0.25}$B$_6$: 
$|g_{\Gamma_5}|=155$~K,
$C^{(0)}_{44} \simeq 8.2 \times 10^{11}$~erg/cm$^3$,~\cite{Suzuki}
and the lattice constant $a=4.13$~\AA.
At absolute zero, the magnitude of quadrupole moments,
induced by the octupole order,
is given by
$|\langle O_{yz \ i} \rangle|=|\langle O_{zx \ i} \rangle|
=|\langle O_{xy \ i} \rangle|=2/3$,
and we obtain
\begin{equation}
  |\epsilon_{yz}|=|\epsilon_{zx}|=|\epsilon_{xy}|=4.6 \times 10^{-5}.
  \label{eq:strain}
\end{equation}

In order to account for the observed lattice contraction
along [111],~\cite{Akatsu,Akatsu2}
we consider two possibilities: (i) $g_{\Gamma_5}<0$
and (ii) $g_{\Gamma_5}>0$. 

\subsection{case \rm{(i)} $g_{\Gamma_5}<0$}
In case (i),
we assume that
a small stress which accompanies the measurement
breaks the equivalence of four octupole domains, 
and chooses a domain
for which the contraction becomes maximum along [111] with $g_{\Gamma_5}<0$.
In this case,
the domain with $\langle O_{yz \ i} \rangle=\langle O_{zx \ i} \rangle
=\langle O_{xy \ i} \rangle= -2/3$ is chosen,
and all $\epsilon_{\alpha \beta}$ are negative.
Then we obtain
\begin{equation}
  \Delta l/l=\frac{1}{3}\epsilon_{\text{B}}
  +\frac{2}{3} (\epsilon_{yz}+\epsilon_{zx}+\epsilon_{xy})
  =\frac{1}{3}\epsilon_{\text{B}}+2\epsilon_{xy},
  \label{eq:caseI}
\end{equation}
along [111],
where $\epsilon_{\text{B}}$ is the volume strain
which is not included in our model.
Experimentally,
the value of the volume strain in Ce$_{0.7}$La$_{0.3}$B$_6$
was obtained to be
$\epsilon_{\text{B}} = 8 \times 10^{-6}$ at 1.3~K.~\cite{Akatsu2}
Note that along $[\bar{1}11] ,[1\bar{1}1]$
and $[11\bar{1}]$ directions, the lattice should expand by 
\begin{equation}
  \Delta l/l
  =\frac{1}{3}\epsilon_{\text{B}}+\frac{2}{3}|\epsilon_{xy}|.
\end{equation}

\subsection{case \rm{(ii)} $g_{\Gamma_5}>0$}
In case (ii),
a positive stress along [111] may favor a domain
which contracts along this direction. 
In this case, two of $\epsilon_{\alpha \beta}$ are negative
and the other is positive.
Then the contraction along [111] is given by,
\begin{equation}
  \Delta l/l
  =\frac{1}{3}\epsilon_{\text{B}}
  +\frac{2}{3} (\epsilon_{yz}+\epsilon_{zx}+\epsilon_{xy})
  =\frac{1}{3}\epsilon_{\text{B}}
  -\frac{2}{3}|\epsilon_{xy}|.
  \label{eq:caseII}
\end{equation}
The crystal expands along another direction.
In the single domain with
$\langle O_{yz} \rangle =\langle O_{zx} \rangle =-\langle O_{xy} \rangle$,
for example, 
an expansion along $[11\bar{1}]$ should be present,
and it is given by
\begin{equation}
  \Delta l/l
  =\frac{1}{3}\epsilon_{\text{B}}
  +\frac{2}{3} (-\epsilon_{yz}-\epsilon_{zx}+\epsilon_{xy})
  =\frac{1}{3}\epsilon_{\text{B}}
  +2|\epsilon_{xy}|.
\end{equation}

\subsection{Comparison with the Experiment}
According to an experiment,~\cite{Akatsu2} the shear strain
on the assumption of the trigonal symmetry around [111],
i.e., case (i),
is derived as
$\epsilon_{\alpha\beta}=-4 \times 10^{-6}$ without a magnetic field
at 1.3~K.
Assuming that the phase IV is stable down to zero temperature, 
$\epsilon_{\alpha\beta}$ is extrapolated to
$-6 \times 10^{-6}$ to $-10 \times 10^{-6}$ at absolute zero.
The absolute value is by an order of magnitude smaller than
our estimate in eq.~(\ref{eq:strain}).
Thus the observed reduction of 
$|\langle O_{\alpha\beta} \rangle|$ 
should come from quantum fluctuations neglected here.

Another possibility is that case (ii) is realized.
In this case, the value of the strain
becomes three times larger, i.e.,
$|\epsilon_{\alpha \beta}| = (2 $--$ 3)\times 10^{-5}$,
when deduced from the same experimental result.~\cite{Akatsu2}
This value is comparable to that given in eq.~(\ref{eq:strain}).
As mentioned in {\S}~\ref{sec:M},
a state where one of $\langle O_{\alpha\beta} \rangle$'s
has a sign different from the others is stabilized
under $\mib{H} \parallel$ [111] without an external stress.
Experimentally the contraction is enhanced
under $\mib{H} \parallel$ [111].~\cite{Akatsu,Akatsu2}
This feature favors case (ii) in our model.

To determine which case is realized
in the experiment,
it is desirable to measure the lattice distortion
along the [110] direction.
The lattice distortion along [110] is given by
$\Delta l/l=(\epsilon_{\text{B}}/3)+\epsilon_{xy}$.
Thus, in case (i),
our model predicts a smaller magnitude of lattice contraction 
along the [110] direction as compared with
the contraction (\ref{eq:caseI}) along the [111] direction.
On the other hand, in case (ii),
the magnitude of lattice contraction
along [110] is expected to be larger
than that of the contraction (\ref{eq:caseII}) along [111].

\subsection{NdB$_6$}
For comparison,
we estimate in the same manner the magnitude of the lattice distortion
of NdB$_6$.
This material undergoes an A-type antiferromagnetic order
at $T_{\text{N}} \simeq 8$~K~\cite{McCarthy}
accompanying a lattice distortion along the [001] direction.~\cite{Sera}
This lattice distortion indicates a $O^0_2$ ferro-quadrupole order.
We estimate the lattice distortion
in the $O^0_2$ ferro-quadrupole-ordered state
by using the experimental values:
$|g_{\Gamma_3}|=220$~K,
$(C_{11}-C_{12})/2
\simeq 20 \times 10^{11}$~erg/cm$^3$,~\cite{Tamaki}
$a=4.12$~\AA,
and the Lea-Leask-Wolf parameter $x_{\text{LLW}}=-0.82$.~\cite{Pofahl}
We assume that $(C^{(0)}_{11}-C^{(0)}_{12})/2$
is almost the same as the observed
$(C_{11}-C_{12})/2$.
In a magnetic field $\mib{H} \parallel$ [100],
a state with the antiferromagnetic moment perpendicular to [100]
is stabilized.
For a domain where the magnetic moment is along [001] or [010],
we obtain
\begin{equation}
  |\Delta l|/l=2.8 \times 10^{-4},
\end{equation}
along $\mib{H}$, on the assumption $\epsilon_{\text{B}}=0$. 
This value is favorably comparable to the experimental one
$\Delta l/l \simeq -2 \times 10^{-4}$ at 2~K and $H=2.1$~T.~\cite{Sera}
The distortion in NdB$_6$ is by an order of magnitude larger 
than those 
in eqs.~(\ref{eq:caseI}) and (\ref{eq:caseII}).
The reason is that the quadrupole moment
$|\langle O^0_2 \rangle| =4.5$ in NdB$_6$
is much larger than the corresponding value
$|\langle O_{xy} \rangle|=2/3$
in the octupolar state of Ce$_x$La$_{1-x}$B$_6$,
while the magneto-elastic coupling constants are of the same order.

\section{Internal Field}\label{sec:field}
In this section,
we discuss
the internal magnetic field associated with the octupole order.
The $\mu$SR time spectra in the phase IV
consist of a Gaussian component
and an exponential component.~\cite{Takagiwa}
The observation of a Gaussian relaxation
indicates that
internal fields are randomly distributed
and/or fluctuating.
The internal field deduced from the Gaussian relaxation
is of the order of 0.1~T.
We discuss whether
the octupole moment can be the origin
of the Gaussian relaxation.
In the $\mu$SR measurement,
$\mu^+$ locates at $(a/2,0,0)$ and equivalent sites~\cite{Amato,Kadono,Schenck}
with a Ce ion chosen as the origin.
As a reference the internal field from
a Bohr magneton $\mu_B$ is estimated to be
\begin{equation}
  H_{\text{dipole}} = \frac{1}{(a/2)^3} \mu_{\text{B}} \simeq 0.1 \text{T},
\end{equation}
with $a/2\simeq 2$~\AA.
The internal field from an octupole with the size $r$ is estimated to be
\begin{equation}
  H_{\text{octupole}} = \frac{r^2}{(a/2)^5} \mu_{\text{B}}
  \simeq 0.01 \text{T},
  \label{eq:octupole-field}
\end{equation}
with $r \simeq a_B=0.53$~\AA.
For a more accurate estimate, we consider the multipole expansion
of the vector potential from local electrons,
which is given by~\cite{Schwartz}
\begin{equation}
  \mib{A}(\mib{r})=\sum_{k,m}
  \frac{-\text{i}}{k} r^{-(k+1)}
  \left(\mib{l}C^{(k)}_{m}(\theta,\phi)\right)M^{m}_{k},
\end{equation}
where
$\mib{l}$ is the orbital angular momentum operator,
$C^{(k)}_{m}(\theta,\phi)$
is $\sqrt{4\pi/(2k+1)}$ times
the spherical harmonics $Y_{k m}(\theta,\phi)$,
and $M^{m}_{k}$ is the magnetic multipole moment.
The multipole moment
is determined by 
the wave function $\psi_i(\mib{r})$ of the $i$-th electron,
and the orbital and spin angular momentum operators
$\mib{l}_i$ and $\mib{s}_i$.
Namely we have
\begin{equation}
  \begin{split}
    M^{m}_{k}=&\mu_{\text{B}}
    \sum_i \int \text{d}\mib{r}_i
    \psi^*_i(\mib{r}_i)
    \left(\mib{\nabla}_i r^k_i C^{(k) *}_{m}(\theta_i,\phi_i)\right)\\
    &\cdot 
    \left(\frac{2}{k+1}\mib{l}_i+2\mib{s}_i\right)
    \psi_i(\mib{r}_i).
  \end{split}
  \label{eq:Mmu}
\end{equation}
Equation~(\ref{eq:Mmu}) is evaluated
with the use of the operator equivalents method.~\cite{Inui}
For our purpose, it is sufficient to
consider only octupole moments.
Then we obtain $M^{m}_3$ through calculation of the reduced matrix element
of the third-rank tensor.
The result for one-electron states with $J=5/2$, $L=3$, $S=1/2$ is given by
\begin{equation}
  M^{m}_3=-\frac{2}{35}\mu_{\text{B}}
  \langle r^2 \rangle
  \langle J^{(3)}_{m} \rangle,
\end{equation}
where
the third-rank tensor operators
$J^{(3)}_{m}$ are defined by
\begin{align}
  J^{(3)}_3 &=-\frac{\sqrt{5}}{4}(J_x+\text{i}J_y)^3,\\
  J^{(3)}_{m-1} &= \frac{1}{\sqrt{(3+m)(4-m)}}
  [J_x-\text{i}J_y,J^{(3)}_m].
\end{align}
We introduce linear combinations: 
\begin{align}
  M^{2u}_3&=
  \frac{1}{\sqrt{2}\text{i}}
  [M^{+2}_3-M^{-2}_3],\label{eq:M2u}\\
  M^{4u x}_3&=
  \frac{1}{4}
  [-\sqrt{5}(M^{+3}_3-M^{-3}_3)
  +\sqrt{3}(M^{+1}_3-M^{-1}_3)],\\
  M^{4u y}_3&=
  \frac{1}{4\text{i}}
  [\sqrt{5}(M^{+3}_3+M^{-3}_3)
  +\sqrt{3}(M^{+1}_3+M^{-1}_3)],\\
  M^{4u z}_3&=
  M^{0}_3,\label{eq:M4uz}\\
  M^{5u x}_3&=
  \frac{1}{4}
  [\sqrt{3}(M^{+3}_3-M^{-3}_3)
  +\sqrt{5}(M^{+1}_3-M^{-1}_3)],\\
  M^{5u y}_3&=
  \frac{1}{4\text{i}}
  [\sqrt{3}(M^{+3}_3+M^{-3}_3)
  -\sqrt{5}(M^{+1}_3+M^{-1}_3)],\\
  M^{5u z}_3&=
  \frac{1}{\sqrt{2}}
  [M^{+2}_3+M^{-2}_3].
\end{align}
We then obtain $\mib{H}=\mib{\nabla} \times \mib{A}$ as
\begin{equation}
  \begin{split}
    H_z(\mib{r})=\frac{1}{36r^5}
    \bigl[
    &6\sqrt{15}O_{xy}(\mib{r})
    \left(7\sqrt{3}O^{0}_{2}(\mib{r})+4\right)M^{2u}_3\\
    +&15O_{zx}(\mib{r})
    \left(7\sqrt{3}O^{0x}_{2}(\mib{r})-2\right)M^{4u x}_3\\
    +&15O_{yz}(\mib{r})
    \left(7\sqrt{3}O^{0y}_{2}(\mib{r})-2\right)M^{4u y}_3\\
    +&2
    \left(105O^{0}_{2}(\mib{r})O^{0}_{2}(\mib{r})
      -20\sqrt{3}O^{0}_{2}(\mib{r})-28\right)M^{4u z}_3\\
    +&9\sqrt{15}O_{zx}(\mib{r})
    \left(7O^{2x}_{2}(\mib{r})+2\right)M^{5u x}_3\\
    +&9\sqrt{15}O_{yz}(\mib{r})
    \left(7O^{2y}_{2}(\mib{r})-2\right)M^{5u y}_3\\
    +&6\sqrt{15}O^{2}_{2}(\mib{r})
    \left(7\sqrt{3}O^{0}_{2}(\mib{r})+4\right)M^{5u z}_3
    \bigr],
  \end{split}
\end{equation}
where
\begin{align}
  O^{0x}_2(\mib{r}) &= (2x^2-y^2-z^2)/(\sqrt{3}r^2),\\
  O^{0y}_2(\mib{r}) &= (2y^2-z^2-x^2)/(\sqrt{3}r^2),\\
  O^{2x}_2(\mib{r}) &= (y^2-z^2)/r^2,\\
  O^{2y}_2(\mib{r}) &= (z^2-x^2)/r^2,
\end{align}
and the other $O(\mib{r})$'s are
obtained by replacing
$J_\alpha$ in eqs.~(\ref{eq:O20})--(\ref{eq:Obc})
with $\alpha/r$.
The other components $H_x(\mib{r})$ and $H_y(\mib{r})$ are obtained 
by exchanging $x,y,z$ cyclically from above.
In the $\Gamma_{5u}$-ordered state,
we obtain
$M^{5u x}_3=M^{5u y}_3=M^{5u z}_3 \ne 0$,
and eqs.~(\ref{eq:M2u})--(\ref{eq:M4uz}) are zero.

To estimate the internal magnetic field,
we use the value
$\langle r^2 \rangle =1.298$ in atomic unit,
which was obtained
with a relativistic Dirac-Fock calculation
by Freeman and Desclaux.~\cite{Freeman}
Then we obtain about 40~G
as the internal field
at $(a/2,0,0)$ from a Ce ion.
This value is by an order of magnitude smaller than that derived
by the $\mu$SR measurement.~\cite{Takagiwa}
Thus the static octupole moment alone cannot account for
the Gaussian relaxation of the $\mu$SR spectra.
In a future work, we plan to study in more detail fluctuations
in the octupole-ordered state.
If the origin of the $\mu$SR spectra
is not an effect of fluctuations,
we may have to consider a disturbance of the ordered state
by muons.

We mention that Paix\~{a}o \textit{et al.} have  proposed
for the ordered state of NpO$_2$
that a triple-$\mib{q}$ $\Gamma_{5u}$-type octupole ordering is realized
and a triple-$\mib{q}$ $\Gamma_{5g}$ quadrupole
moment is induced.~\cite{Paixao}
From the observed 
muon spin precession frequency of $\omega=7$~MHz~\cite{Kopmann}
and the muon gyromagnetic ratio $\gamma_{\mu}=2\pi \times 13.55$~MHz/kG at 8~K,
we estimate
the internal magnetic field as
$H=\omega/\gamma_{\mu} \simeq 500$~G.~\cite{note}
This internal field 
is smaller than that of antiferromagnetic UO$_2$ by an order of magnitude.
Thus the observed field in NpO$_2$
may be interpreted as coming from octupole moments.
However, the internal field in NpO$_2$ 
is still much larger than that of eq.~(\ref{eq:octupole-field}).
Provided the muon is really probing the intrinsic internal field
of the systems,
the muon stopping site should be about 1~{\AA} from a Np ion.

\section{Pressure Induced Moment}\label{sec:pressure}
\begin{figure}[t]
  \includegraphics[width=8cm]{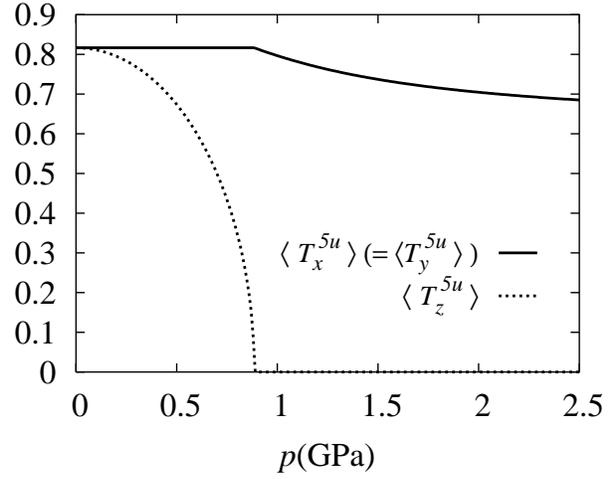}
  \caption{Pressure dependence of
    the antiferro-octupole moment
    $\langle T^{5u}_x \rangle$
    $(=\langle T^{5u}_y \rangle)$
    and  $\langle T^{5u}_z \rangle$.}
  \label{figure:T-p}
\end{figure}

\begin{figure}[t]
  \includegraphics[width=8cm]{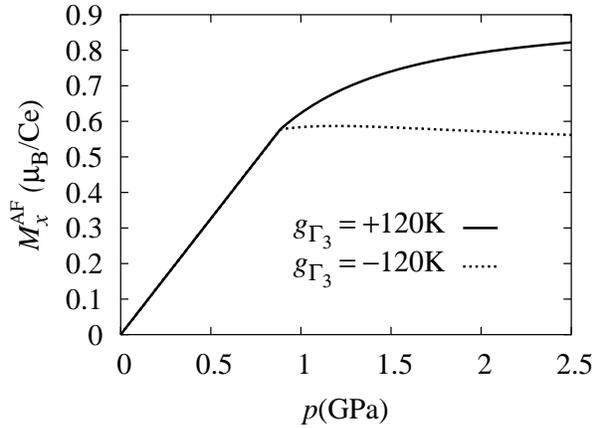}
  \caption{Pressure dependence of the antiferromagnetic
    moment $M^{\text{AF}}_x$ ($=M^{\text{AF}}_y$).
    $M^{\text{AF}}_z=0$.}
  \label{figure:M-p}
\end{figure}

It has been discussed that 
antiferromagnetism should be induced in an antiferro-octupolar phase
by uniaxial stress.~\cite{Kuramoto}
We apply this idea to the case of the $\Gamma_{5u}$ order.
The uniaxial pressure $p$ along the [001] direction accompanies 
the $\Gamma_3$ strain 
\begin{equation}
  \epsilon_{u}=
  \frac{1}{\sqrt{3}}(2\epsilon_{zz}-\epsilon_{xx}-\epsilon_{yy})
  =\frac{p}{\sqrt{3}(C_{11}-C_{12})/2}.
\end{equation}
We solve the mean field equation with the finite
strain $\epsilon_u$
and the corresponding quadrupole-strain interaction $g_{\Gamma_3}$
(see, eq.~(\ref{eq:Eela})).
The antiferromagnetic moment is induced in the $xy$-plane
since $\langle T^{5u}_x \rangle \neq 0$
together with $\langle J^2_y-J^2_z \rangle\neq 0$ gives
$\langle J_x \rangle\neq 0$.
The direction of the antiferromagnetic moment
is along [110] or $[1\bar{1}0]$.
In Figs.~\ref{figure:T-p} and \ref{figure:M-p},
we show the pressure dependence
of the octupole and magnetic moments at absolute zero
with experimental values:
$(C_{11}-C_{12})/2 \simeq 20.4 \times 10^{11}$~erg/cm$^3$
and $|g_{\Gamma_3}|=120$~K.~\cite{Suzuki}
The mean field solution for $\langle T^{5u}_z \rangle$ becomes zero
for $p \geq 0.9$~GPa.
This implies that both the octupole and dipole moments lie in the $xy$-plane.
With $p=1$~GPa, the magnetic moment
is estimated to be 0.88$\mu_{\text{B}}$ (0.82$\mu_{\text{B}}$),
when $g_{\Gamma_3}$ is positive (negative).
This value should actually be reduced
by quantum fluctuations neglected here.
However, it is likely that this value remains within the experimental range.

\section{Summary}
To summarize,
we have proposed that
the $\Gamma_{5u}$ octupole moment
is a plausible candidate for the order parameter of the phase IV
in Ce$_x$La$_{1-x}$B$_6$.
This octupole ordered state is consistent with
the lattice distortion along the [111] direction,~\cite{Akatsu,Akatsu2}
the softening of the elastic constant $C_{44}$,~\cite{Suzuki}
the broken time reversal symmetry,~\cite{Magishi,Takagiwa}
the absence of dipole moment~\cite{Iwasa},
and the cusp in the magnetization at $T_{\text{I-IV}}$.~\cite{Tayama}
Especially,
the sudden softening of $C_{44}$
and the cusp in the magnetization at $T_{\text{I-IV}}$
are difficult to be explained by a pure quadrupole order,
but are reproduced by the present octupole ordering model.

Many properties of the phase IV
are consistent with the octupole order,
but the internal field estimated by our mean field theory
is much smaller than that suggested by
the $\mu$SR experiment.~\cite{Takagiwa}
We recall that the NMR spectra become broad in the phase IV,~\cite{Magishi}
and the spectra cannot be explained
by the static and staggered octupole moments, either.
Hence some dynamical or 
quantum fluctuation effects may be playing a significant role. 
Clarifying dynamical aspects and identifying the periodicity in
the phase IV are the most challenging open problems to be addressed
in the near future.

We have also shown that antiferromagnetic moment
lying in the $xy$-plane is induced
under uniaxial pressure along the $z$ direction.
The estimated value of the antiferromagnetic moment
is of the order of $1\mu_{\text{B}}$, and we wait for 
experimental efforts to detect the moment.

\section*{Acknowledgements}
We thank Prof. A. Schenck
for pointing out our mistake
in the estimation of the internal field in NpO$_2$.
This work was supported partly by
Special Coordination Funds for Promoting Science and
Technology, and by the NEDO international collaboration program
``New boride materials''.


\begin{thebibliography}{99}
\bibitem{Santini}
  P. Santini and G. Amoretti:
  Phys. Rev. Lett. \textbf{85} (2000) 2188.
\bibitem{Paixao}
  J. A. Paix\~{a}o, C. Detlefs, M. J. Longfield, R. Caciuffo,
  P. Santini, N. Bernhoeft, J. Rebizant and G. H. Lander:
  Phys. Rev. Lett. \textbf{89} (2002) 187202.
\bibitem{Caciuffo}
  R. Caciuffo, J. A. Paix\~{a}o, C. Detlefs, M. J. Longfield,
  P. Santini, N. Bernhoeft, J. Rebizant and G. H. Lander:
  J. Phys.: Condens. Matter \textbf{15} (2003) S2287.
\bibitem{Kiss}
  A. Kiss and P. Fazekas:
  cond-mat/0306215.
  
%
\bibitem{Kuramoto}
  Y. Kuramoto and H. Kusunose:
 J. Phys. Soc. Jpn. \textbf{69} (2000) 671.
\bibitem{Kusunose}
  H. Kusunose and Y. Kuramoto:
  J. Phys. Soc. Jpn. \textbf{70} (2001) 1751.

%
\bibitem{Shiina}
  R. Siina, H. Shiba and P. Thalmeier:
  J. Phys. Soc. Jpn. \textbf{66} (1997) 1741.
\bibitem{Sakai}
  O. Sakai, R. Siina, H. Shiba and P. Thalmeier:
  J. Phys. Soc. Jpn. \textbf{66} (1997) 3005.
\bibitem{Shiina2}
  R. Siina, O. Sakai, H. Shiba and P. Thalmeier:
  J. Phys. Soc. Jpn. \textbf{67} (1998) 941.

\bibitem{Sakakibara2}
  T. Sakakibara, T. Tayama, H. Amitsuka,
  K. Tenya, S. Kunii, T. Suzuki and A. Ochiai:
  Physica B \textbf{230-232} (1997) 307.

\bibitem{Suzuki}
  O. Suzuki, T. Goto, S. Nakamura, T. Matsumura and S. Kunii:
  J. Phys. Soc. Jpn. \textbf{67} (1998) 4243.

\bibitem{Tayama}
  T. Tayama, T. Sakakibara, K. Tenya, H. Amitsuka and S. Kunii:
  J. Phys. Soc. Jpn. \textbf{66} (1997) 2268.

\bibitem{Akatsu}
  M. Akatsu, O. Suzuki, Y. Nemoto, T. Goto,
  S. Nakamura and S. Kunii:
  \textit{Proc. Int. Conf. Strongly Correlated Electrons with
  Orbital Degrees of Freedom (ORBITAL2001)},
  J. Phys. Soc. Jpn. \textbf{71} (2002) Suppl., p. 115.
\bibitem{Akatsu2}
  M. Akatsu, T. Goto, Y. Nemoto, O. Suzuki,
  S. Nakamura and S. Kunii:
  J. Phys. Soc. Jpn. \textbf{72} (2003) 205.

\bibitem{Furuno}
  T. Furuno, N. Sato, S. Kunii, T. Kasuya and W. Sasaki:
  J. Phys. Soc. Jpn. \textbf{54} (1985) 1899.

\bibitem{Magishi}
  K. Magishi, M. Kawakami, T. Saito, K. Koyama,
  K. Mizuno and S. Kunii:
  Z. Naturforsch. \textbf{57 a} (2002) 441.

\bibitem{Takagiwa}
  H. Takagiwa, K. Ohishi, J. Akimitsu, W. Higemoto,
  R. Kadono, M. Sera and S. Kunii:
  J. Phys. Soc. Jpn. \textbf{71} (2002) 31.

\bibitem{Iwasa}
  K. Iwasa, K. Kuwahara, M. Kohgi, P. Fischer, A. D\"{o}nni,
  L. Keller, T. C. Hansen, S. Kunii, N. Metoki, Y. Koike
  and K. Ohoyama:
  Physica B \textbf{329-333} (2003) 582.
  
\bibitem{Sakakibara}
  T. Sakakibara, K. Tenya, M. Yokoyama, H. Amitsuka
  and S. Kunii:
  \textit{Proc. Int. Conf. Strongly Correlated Electrons with
  Orbital Degrees of Freedom (ORBITAL2001)},
  J. Phys. Soc. Jpn. \textbf{71} (2002) Suppl., p. 48.
\bibitem{Sakakibara3}
  T. Sakakibara, T. Tayama, K. Tenya, M. Yokoyama, H. Amitsuka,
  D. Aoki, Y. \={O}nuki, Z. Kletowski and S. Kunii:
  J. Phys. Chem. Solids \textbf{63} (2002) 1147.
\bibitem{Morie}
  T. Morie, T. Onimaru, T. Tayama, T. Sakakibara and S. Kunii:
  presented at the autumn meeting of the Phys. Soc. Jpn. (2003).
  
%
\bibitem{Kubo}
  K. Kubo and Y. Kuramoto:
  J. Phys. Soc. Jpn. \textbf{72} (2003) 1859.
  
\bibitem{Zirngiebl}
  E. Zirngiebl, B. Hillebrands, S. Blumenr\"{o}der,
  G. G\"{u}ntherodt, M. Loewenhaupt,
  J. M. Carpenter, K. Winzer and Z. Fisk:
  Phys. Rev. B \textbf{30} (1984) 4052.
\bibitem{Luthi}
  B. L\"{u}thi, S. Blumenr\"{o}der, B. Hillebrands,
  E. Zirngiebl, G. G\"{u}ntherodt and K. Winzer:
  Z. Phys. B \textbf{58} (1984) 31.
\bibitem{Sato}
  N. Sato, S. Kunii, I. Oguro,
  T. Komatsubara and T. Kasuya:
  J. Phys. Soc. Jpn. \textbf{53} (1984) 3967.

\bibitem{Morin}
  P. Morin, A. Waintal and B. L\"{u}thi:
  Phys. Rev. B \textbf{14} (1976) 2972.
  
\bibitem{Callen}
  See, for example, B. Callen:
  \textit{THERMODYNAMICS}
  (John Wiley \& Sons, 1960)

\bibitem{McCarthy}
  C. M. McCarthy and C. W. Tompson:
  J. Phys. Chem. Solids \textbf{41} (1980) 1319.
  
\bibitem{Sera}
  M. Sera, S. Itabashi and S. Kunii:
  J. Phys. Soc. Jpn. \textbf{66} (1997) 548.

\bibitem{Tamaki}
  A. Tamaki, T. Goto, M. Yoshizawa, T. Fujimura,
  S. Kunii and T. Kasuya:
  J. Magn. Magn. Mater. \textbf{52} (1985) 257.

\bibitem{Pofahl}
  G. Pofahl, E. Zirngiebl, S. Blumenr\"{o}der, H. Brenten,
  G. G\"{u}ntherodt and K. Winzer:
  Z. Phys. B \textbf{66} (1987) 339.

\bibitem{Amato}
  A. Amato, R. Feyerherm, F. N. Gygax and A. Schenck:
  Hyperfine Interact. \textbf{104} (1997) 115.
\bibitem{Kadono}
  R. Kadono, W. Higemoto, A. Koda, K. Kakuta,
  K. Ohishi, H. Takagiwa, T. Yokoo and J. Akimitsu:
  J. Phys. Soc. Jpn. \textbf{69} (2000) 3189.
\bibitem{Schenck}
  A. Schenck, F. N. Gygax and S. Kunii:
  Phys. Rev. Lett. \textbf{89} (2002) 037201.

\bibitem{Schwartz}
  C. Schwartz:
  Phys. Rev. \textbf{97} (1955) 380.

\bibitem{Inui}
  See, for example,
  T. Inui, Y. Tanabe and Y. Onodera:
  \textit{Group Theory and Its Applications
    in Physics}
  (Springer-Verlag, Berlin, 1996) 2nd ed.

\bibitem{Freeman}
  A. J. Freeman and J. P. Desclaux:
  J. Magn. Magn. Mater. \textbf{12} (1979) 11.
  
\bibitem{Kopmann}
  W. Kopmann, F. J. Litterst, H.-H. Klau{\ss}, M. Hillberg,
  W. Wagener, G. M. Kalvius, E. Schreier, F. J. Burghart,
  J. Rebizant and G. H. Lander:
  J. Alloys Compounds \textbf{271-273} (1998) 463.

\bibitem{note}
  In our previous paper,~\cite{Kubo}
  we have mistaken the estimation of the internal field in NpO$_2$
  by the factor of $2\pi$.

\end{thebibliography}
\end{document}